\documentclass[review,preprint]{revtex4}
\usepackage[dvips]{graphicx}
\usepackage{t1enc,amsmath}
\usepackage{amsfonts}
\usepackage{amssymb}
\begin{document}

\title{Resonant bowtie aperture nano-antenna for the control of optical nanocavities resonance}
\author{Fadi I. Baida and Thierry Grosjean}
\affiliation{D\'epartement d'Optique P.M. Duffieux, Institut FEMTO-ST, UMR 6174 CNRS \\
 Universit\'{e} de Franche--Comt\'{e}, 25030 Besan\c{c}on Cedex, France}\email{fadi.baida@femto-st.fr}

\begin{abstract}
Scanning Near-field Optical Microscopy (SNOM) has been successful in
finely tuning the optical properties of photonic crystal (PC)
nanocavities. The SNOM nanoprobes proposed so far allowed for either
redshifting or blueshifting the resonance peak of the PC structures.
In this Letter, we theoretically demonstrate the possibility of
redshifting (up to $+0.65$~nm) and blueshifting (up to $-5$~nm) PC cavity
resonance with a single SNOM probe. This probe is obtained by
opening a bowtie-aperture nano-antenna (BNA) at the apex of a
metal-coated tip. This double-way PC tunability is the result of a
competition between the effects of the BNA resonance (induced electric dipole leading to a redshift) and the metal-coated tip (induced magnetic dipole giving rise to a blueshift) onto the PC
mode volume. The sign of the spectral shift is modified by simply
controlling the tip-to-PC distance. This study opens the way to the full
postproduction control of the resonance wavelength of high quality
factor optical cavities.
\end{abstract}

\maketitle

\section{Introduction}
It is well known that for a given optical cavity, the
resonance properties depend on the volume allocated to the excited
mode. This property generated several recent studies that deal with
the resonance wavelength (RW) control especially in the case of high
quality-factor resonators such as photonic crystal cavities. Modifications of
this volume can be achieved through different physical processes.
Thus, optical \cite{belotti:oe08,maksymov:oe11}, electrical
\cite{fushman:apl07,majumdar:nl13}, chemical
\cite{elkallassi:josab07,cai:apl13} or mechanical
\cite{vignolini:apl10,descharmes:prl13} effects were exploited to
shift the RW by slightly modifying the cavity resonance properties.
Plasmonic \cite{barth:nl10} and, more widely, any Fano-type
resonances are also ideal candidates for such control.

In all cases, the perturbation must be small enough not to cancel the
cavity resonance. Basically, the cavity mode volume can be modified
by immersing a small dielectric element within the evanescent field
of the cavity
\cite{koenderink:prl05,marki:oe06,hopman:oe06,lalouat:prb07,legac:oe09}.
Thereby, the effective volume of the latter is increased (light
penetrates inside the dielectric) leading to a red shift of its RW.
Contrarily, if we replace dielectric by metal that inhibits the
light permeation, the electromagnetic field will be mostly squeezed
so that the whole mode volume decreases. The direct consequence is
then a blue shift of the RW \cite{maksymov:oe11,grosjean:apl12}.
Nonetheless, if the metallic structure exhibits localized resonance
(almost with dipolar behavior), redshifts may occur after complete
compensation of the mode squeezing due to the presence of the metal as
already demonstrated in reference \cite{barth:nl10}. Generally, for
these effects which are opposed, two cases are possible: either they
are compensated, either one of them is dominant such that we can not
discern them \cite{burresi:prl10}. In all cases, the overall
frequency shift is so small ($\Delta\lambda/\lambda\simeq10^{-4}$)
that resonators of very large Q-factors are needed
\cite{mujumdar:oe07} for experimental evidence. Thus, the question
is how to control the contribution of each effect in order to
actively control the RW? A basic idea consists of the design of an
element that allows the two above cited effects and also permits
adjusting the influence of each one through its interaction with the
optical cavity. Nevertheless, the interaction of the cavity with
this element should be as weak as possible to not completely break
down the cavity resonance.

Appropriately, to highlight the existence of these two effects within a single configuration, some authors proposed a metal coated SNOM tip with small aperture at its apex in order to control the resonance wavelength of an optical cavity \cite{burresi:prl10}. Such a perturbation element combines, at least, an induced electric dipole together with a magnetic one. Unfortunately, the cavity-tip interaction was dominated by the magnetic contribution due to the large metal volume of the tip apex that is involved in the interaction. Nevertheless, theoretical calculations based on FDTD numerical simulations, showed that it is possible to get both red (very weak shift of only $30pm$) and blue shifts (almost $2.8${\AA}) with the same SNOM tip by placing its apex at specific positions in the vicinity of the cavity. By the way, the discrimination between the two effects is done spatially by exciting selectively the electric or the magnetic dipolar response of the tip through the knowledge of the electromagnetic field near the cavity.\\
As it is well-known, the magnetic effect, that is induced by the
longitudinal component of the magnetic field (here along the tip
axis) and leads to a blueshift, can be seen as a consequence of the
cavity mode squeezing by the metallic apex of the tip. The
electric effect of the tip apex is induced by the transversal
component of the electric field around the tip apex and leads to a
redshift of the resonance wavelength that can be seen to be
due to an increase of the cavity mode volume. Based on Refs.
\cite{koenderink:prl05,vignolini:prl10,burresi:prl10}, the
contribution of the electric and magnetic dipoles to the wavelength
shift can be written as:

\begin{equation}
\frac{\Delta\lambda}{\lambda}=\left[\frac{\Delta\lambda}{\lambda}\right]^e+\left[\frac{\Delta\lambda}{\lambda}\right]^m=({\alpha^e}|E_{//}|^2+{\alpha^m}|B_z|^2)\frac{e^{-2d/L_0}}{W_0}\\
\label{relativeshift}
\end{equation}

Where $\alpha^{e}$ and $\alpha^{m}$ are electric and
magnetic polarizabilities of the BNAT respectively, $E_{//}$ is the transverse
component of the electric field within the unperturbed cavity,
$B_z$ is the magnetic longitudinal component at the top of the
unperturbed cavity and $W_0$ is the total stored energy in the PC cavity at resonance. The decay term $e^{-2d/L_0}$ is the same for
both magnetic and electric field and allows the determination of the
fields at the tip apex position (at a distance $d$ from the cavity)
through the knowledge of the decay length  $L_0$ . The $\alpha^e$
and $\alpha^m$ terms have an opposite sign and both depend on the tip apex (the nature and the volume of metal that is involved by the tip-PC interaction for the magnetic term $\alpha^m$ and the size and shape of the aperture where the electric effect takes place for the electric term $\alpha^e$). Consistently with
the results of ref. \cite{burresi:prl10} and with recent theoretical and
experimental investigations \cite{rotenberg:prl12,rotenberg:prb13}, a predominant
contribution (from 2.5:1 to 10:1) of the magnetic effect was demonstrated compared to the
electric one in the case of circular hole aperture engraved in metallic film.

In order to bypass this prevalence of the magnetic effect, a
perturbation element that exhibits electric resonance would be
necessary to reinforce the electric effect of the tip. More precisely, this resonance must exhibit a small quality factor in
order not to interact destructively with the optical cavity. Gap-based nano-antennas are ideal candidates due to their optical
resonance that gives rise to spatially confined electric field over a broad
spectral range (quality factor of around 3) \cite{eleter:oex14}.

In this paper, we propose an example of such configuration where an
efficient resonator having quite large quality factor, a photonic crystal
cavity, is coupled to a bowtie nano-aperture antenna (BNA)
engraved at the end of a SNOM tip (called BNAT in the following).
Due to the electric resonance of the BNA, the value of the ratio $ \frac{\alpha^e}{W_0}$
is not only enhanced but it will depend on the distance between the tip and the cavity. In fact, when the BNA
resonates, the interaction between the cavity and the BNAT can
significantly decrease the stored energy ($W_0$) in the cavity due to a funneling effect \cite{mivelle:oe14} towards
the BNA that dissipates a significant part of the energy by both
scattering and absorption. In this case, the quasi-static
approximation (a fortiori the depolarization regime \cite{inoue:jap03}) is not
fulfilled because the size of the BNAT apex cannot be considered
larger than its skin depth at resonance due to the light funneling phenomenon toward the BNA. Analytic
expression of $\alpha^e$ can then hardly be derived contrarily to the case
of quasi-static (QS) approximation. In fact, as it will be demonstrated in the
following, the QS approximation that leads to a constant value of
$\alpha$ terms is only fulfilled when the BNA is off-resonance i.e. when the electric effect is canceled (only $\alpha^m$ is constant).
Moreover, at the BNA resonance, the coupling between the two resonators greatly depends
on the distance between them and thus the electromagnetic field above the PC does not undergo a basic evanescent decay as described in Eq. \ref{relativeshift}. Indeed, a critical coupling between the two resonators can take place for specific value of the BNA-to-PC distance \cite{belarouci:oex10,eleter:oex14}. This sensitivity to the distance $d$, which will be presented in section \ref{couplage}, is exploited to control the weight of the tip electric effect with respect to the
magnetic one onto the PC cavity.

\section{The proposed structure}

\subsection{Study of the BNA}
\begin{figure}[ht]
\centering
\includegraphics[width=8cm]{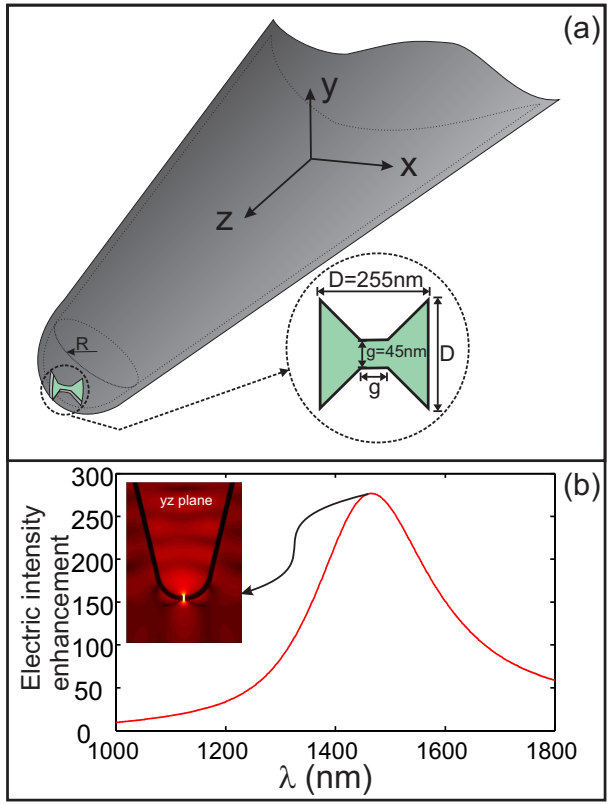}
\caption{(a) Schematic of the BNA engraved at the tip apex of radius
$R$. (b) Enhancement factor of the BNAT measured $15$~nm in front of
its apex and defined as the ratio of the intensity with BNA by the
same intensity calculated without metal coating. The inset of (b)
shows the intensity distribution (logarithmic scale) in the $yOz$
longitudinal plane to point out the light confinement occurring in
the gap zone. } \label{bnaschema}
\end{figure}

The BNA was designed through its geometrical parameters (see figure
\ref{bnaschema}a) to obtain a resonance wavelength close to the one
of the PC cavity. Thus, on the basis of ref.
\cite{mivelle:oe10}, a metal coated SNOM tip with $100~nm$ thick
aluminum layer and an apex radius $R$ is considered to receive the
BNA. The latter is supposed to be centered on the tip axis and has a
gap of $g=45~nm$ and a side width of $D=255~nm$ (see figure
\ref{bnaschema}a). The near-field spectral response of the tip,
calculated by 3D-FDTD home-made code, is presented in figure
\ref{bnaschema}b in addition to the spatial distribution of the
electric field intensity in the $yOz$ longitudinal plane at its
resonance (inset of the same figure). The near-field spectrum given
in figure \ref{bnaschema}b is calculated at $15~nm$ in front of the
BNAT when it is illuminated by the fundamental guided mode of the optical fiber. Normalization of this
near-field signal by the same intensity calculated without the BNA
(uncoated tip) leads to an enhancement factor that reaches $280$. 

As clearly seen from the inset of figure \ref{bnaschema}b, the
resonant BNA acts as a light concentrator. Note that this resonance is only
obtained for the electric field component that is parallel to the
metallic arms of the BNA ($Oy$ direction here). The resulting
polarization sensitivity of the BNA allows distinction between
the two transversal electric field components
\cite{mivelle:oe10,mivelle:oe14} and will be used to discriminate between the magnetic effect of the electric one. Let us emphasize that this
resonance corresponds to the excitation, at its cutoff wavelength,
of the BNA fundamental guided mode that propagates through the metal
thickness \cite{idriss:ol10}. For a given metal, this cutoff
wavelength can be analytically expressed as a function of the BNA
geometrical parameters as demonstrated in ref. \cite{idriss:ol10}
allowing a very simple design of the BNA.

\subsection{Study of the photonic crystal cavity}

\begin{figure}[ht]
\centering
\includegraphics[width=12cm]{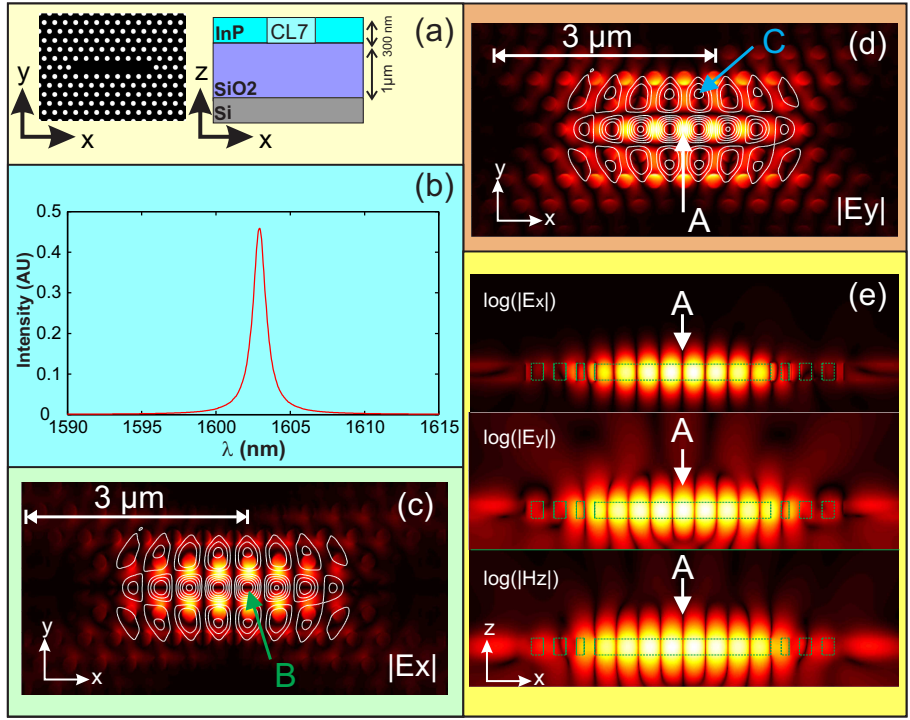}
\caption{Schematic of the proposed CL7 cavity in (a) with vertical
and horizontal cross sections along the structure. The near-field
spectrum, calculated at the center of the cavity, is presented in
(b) where the resonance of the fundamental mode is shown to occur at
$\lambda^R_0=1602.94~nm$ with a quality factor of $Q_{CL7}=1590$. The amplitude of the two transversal electric field
components at this resonance are given in
the $xOy$ plane located $15~nm$ above the InP membrane in (c) and (d). The white lines are iso-amplitude contours of the vertical component of the magnetic field ($H_z$)
(e) Spatial distributions of the two transversal electric field and the longitudinal magnetic field components (logarithmic
scale) at resonance in a vertical plane ($xOz$) passing by the center of the CL7.} \label{cl7}
\end{figure}

Let us now consider the photonic crystal -based optical cavity. It is composed of seven missed
and aligned air holes in a triangular lattice photonic crystal that
was designed to exhibit confined in-plane Bloch modes (period
$a=420~nm$ and hole radius $r=0.25a$)\cite{monat:jap03}. In
addition, the two air holes located at the edges of the cavity are
slightly moved out from the cavity in order to enhance the quality
factor of its fundamental mode \cite{akahane:nature03}. The PC is
made in a InP layer of $300~nm$ thickness deposited on a $1~\mu m$
silica layer separating the InP from silicon substrate (see figure
\ref{cl7}a). The fundamental mode of the cavity cannot be excited by
propagating incident wave since it is spectrally located under the
light line in the dispersion diagram of the PC structure.

Therefore, the excitation of the CL7 cavity is simulated with a
3D-FDTD homemade code by considering a point source (dipole)
emitting over a wide spectral range and positioned in the
vicinity of the cavity so that its orientation and position agree with the excitation of the cavity fundamental mode. The dipole source has a physical meaning
since it can model a quantum emitter coupled to the cavity
\cite{novotnyhecht:book06}. This coupling leads to the excitation of
several cavity modes, thus, we have restricted our spectral study to
a small spectral range centered on the resonance wavelength of the
fundamental mode of the cavity. The near field spectral density of the
CL7 cavity is shown on figure \ref{cl7}b with a RW equal to
$\lambda_0^R=1602.94nm$ and a quality factor estimated to
$Q_{CL7}=1590$. 

3D-FDTD
results giving the spatial distributions, at $15nm$ above the PC cavity, of the transversal electric field
components ($E_x$ and $E_y$) are shown in figures \ref{cl7}c and d while the longitudinal magnetic field component ($H_z$) is given by the white contour plots on the same figures.
On the first hand, one can see
from these two figures that the transversal (x and y) components are
always spatially interleaved so that the $E_x$ component vanishes
when the $E_y$ one is maximum. This specific property is of interest
for the demonstration of the RW tunability in our case. In other
words, when the BNA is placed above the center of the cavity (A position in figure \ref{cl7}c), its
state (on- or off-resonance) can be simply controlled through its
direction: on-resonance if the BNA's metallic arms are along the
$y$-direction and off-resonance when they are along the
$x$-direction, without changing the BNAT spatial position \cite{mivelle:oe14}. This
allows here for differentiating the interaction effect induced by the BNA
resonance (dielectric term) from that of the tip metal coating
(magnetic response). From figure \ref{cl7}e, that shows the spatial distribution of the two
transverse electric field components and the longitudinal magnetic one in the $xOz$ vertical plane, we estimated the value of the decay length
to be $L_0=98~nm$. We verified that, both electric and magnetic
fields have the same decay length.

\section{Study of the coupling}

\label{couplage}
The whole structure (BNAT + CL7) is numerically studied with a
3D-FDTD homemade code that incorporates non-uniform mesh together
with a sub-griding technique in order to faithfully describe the
structure. First, the BNAT is supposed to be centered on the CL7 cavity.
According to figure \ref{cl7}c and d, only the $E_y$ component of
the electric field exists at this location (A), $E_x$ is
equal to zero. Consequently, the BNA metallic arms must be oriented
along the $y$-direction to induce its resonance.

\begin{figure}[ht]
\centering
\includegraphics[width=10cm]{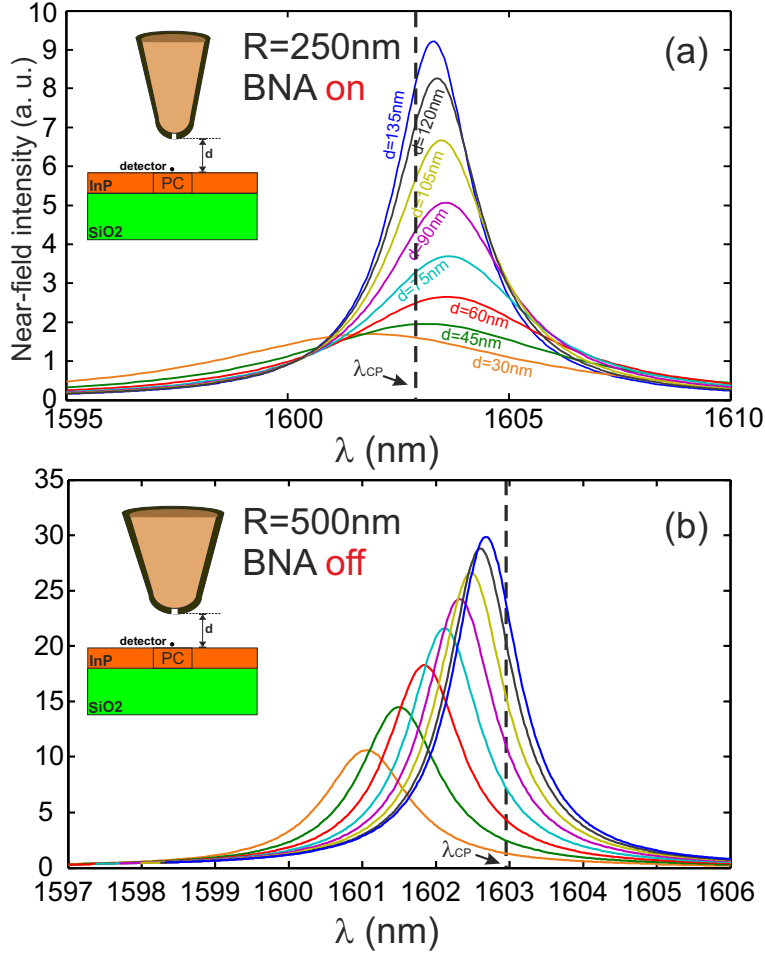}
\caption{Near field spectra calculated $15~nm$ in front of the PhC
for two different BNAT. In (a) the BNA resonates meaning that its
metallic arms are along the $y$-direction. In addition, the tip
radius is fixed to $R=250~nm$ in order to minimize the blue shift
that directly depends on the tip size. Consequently, blueshift only
appears for a small distance $d=30~nm$ while the BNA resonance is
responsible for the redshift because a part of the light penetrates
inside the gap leading to a larger mode volume. In (b) the radius of
the tip is increased to $R=500~nm$ to enhance the mode squeezing and
the BNA is rotated by $90^o$ in order to inhibit its resonance. In
this case, the BNAT almost plays the role of a metallic rod as in
ref. \cite{maksymov:oe11} leading only to a blueshift whatever the
value of the distance $d$ is as experimentally demonstrated in
\cite{grosjean:apl12}.} \label{spectra}
\end{figure}

Optical near-field spectra calculated $15~nm$ above the CL7 are
presented in figure \ref{spectra}a for different values of the
BNAT-to-CL7 distance ($d$) when the BNA is oriented along the $Oy$ direction
(BNA on-resonance) and for a tip apex of $R=250~nm$. We recall that $R$ is the dielectric tip apex radius before the metal coating). One can clearly see form figure \ref{spectra}a the
occurrence of two opposite behaviors: a significant redshift for
large value of $d$ and a small blue one for $d=30~nm$. Thus, two different phenomena are in competition: the metallic tip
effect or magnetic effect that squeezes the PC cavity mode (blue shift) and the BNA
resonance that tends to increase the mode volume by extending light distribution
inside its gap (dielectric resonance). This interpretation is clearly confirmed thanks to the
results of figure \ref{spectra}b obtained for off-resonant BNA and
showing, similarly to reference \cite{grosjean:apl12}, only blue
shifts whatever the distance separating the tip from the cavity.
However, the quality factor of these resonances evolves differently
depending on the BNA state (on or off) as it will be seen in the following.

\begin{figure}[ht]
\centering
\includegraphics[width=8cm]{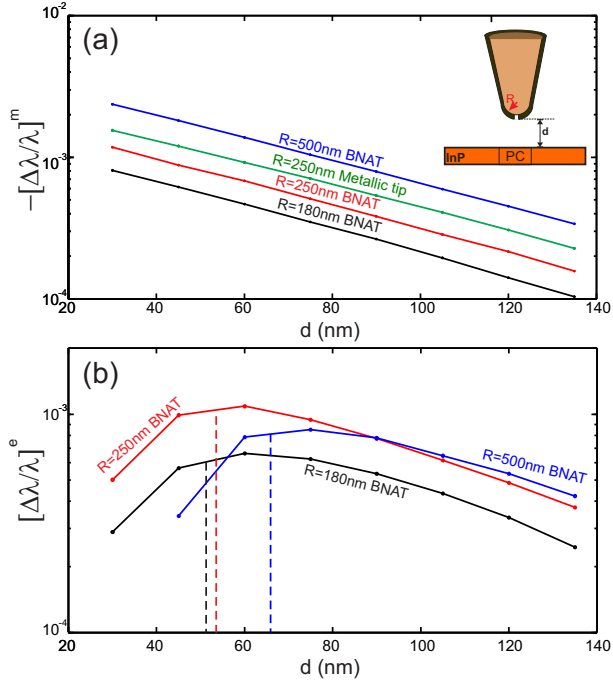}
\caption{Magnetic in (a) and electric in (b) relative shifts of the resonance wavelength of the CL7 fundamental mode as a function of the BNAT-to-CL7 distance $d$. Three different BNAT apex radius are considered ($R=180nm$ in black, $R=250nm$ in red and $R=500nm$ in blue). The magnetic shift induced by an apertureless SNOM tip with radius $R=250nm$ is also presented in green line in (a).} \label{cpcp}
\end{figure}

Second, in order to quantify the magnetic and the electric effects, we have numerically studied the coupling for different values of the tip radius and for both on- and off-resonant BNA. Figure \ref{cpcp}a shows the variations of the relative shift as a function of the BNAT-to-PC distance for off-resonant BNA when the tip is placed in the center of the CL7 cavity (point A indicated by the white arrow in figure \ref{cl7}d). Note that at this location, the longitudinal magnetic field vanishes (see white contours) and the obtained magnetic effect is due to the interaction of the tip sidewalls with the closest lobes of the magnetic field. In this case, the electric term of Eq. \ref{relativeshift} is negligible and only magnetic coupling occurs between the BNAT and the PC (second term of Eq. \ref{relativeshift}). The linear behaviors observed in figure \ref{cpcp}a confirm that the magnetic effect can be derived from a quasi-static (QS) approximation as in Ref. \cite{burresi:prl10} whatever the value of the tip apex radius (from $R=180nm$ to $R=500nm$). 

On the other hand, the slope of the four lines in figure \ref{cpcp}a, which is equal to $\frac{-2}{L_0}$ is in a excellent agreement with the decay length $L_0=98nm$ of the electromagnetic field above the unperturbed cavity. Values of $L_0=108nm$, $105nm$ $104nm$ and $102nm$ are obtained for $R=500nm$, $250nm$, $300nm$ and $180nm$ respectively. Once again, the QS approximation is validated in the case of passive BNAT (off-resonant BNA). Nevertheless, the small increasing in the decay length with respect to the tip radius can be seen as a slight increasing of the longitudinal component of the magnetic field ($B_z$) with respect to the unperturbed cavity due to the coupling with the BNAT.
Nonetheless, we have calculated the cavity energy $W_0$ and determine the effective magnetic polarizability, for the four values of the radius, through Eq. \ref{relativeshift}. The obtained values are : $\alpha^m_{R=180nm}=-8.53\times10^{-21}m^3/\mu_0$, $\alpha^m_{R=250nm}=-12.53\times10^{-21}m^3/\mu_0$, $\alpha^m_{R=300nm}=-15.37\times10^{-21}m^3/\mu_0$ and $\alpha^m_{R=500nm}=-25.96\times10^{-21}m^3/\mu_0$. These values well agree with the value ($\alpha^m_{R=200nm}=-12\times10^{-21}m^3/\mu_0$) obtained in ref. \cite{burresi:prl10} for a cylindrical aperture metal coated SNOM tip with a radius of $200nm$. Note that the magnetic polarizability is directly linked to the metal volume that is involved in the interaction so that the presence of the aperture will slightly decrease its value. We have verified this assumption through the determination of the polarizability of an apertureless metallic tip with radius $R=250nm$ and get a value of $\alpha^m=-17.44\times10^{-21}m^3/\mu_0$ instead of $\alpha^m_{R=250nm}=-12.53\times10^{-21}m^3/\mu_0$ with BNA. By the way, we stress the fact that apertures in SNOM tip are not necessary to induce the magnetic effect; simple metallic nano-rod is more efficient as already demonstrated in ref. \cite{maksymov:oe11}.
\begin{figure}[ht] \centering
\includegraphics[width=10cm]{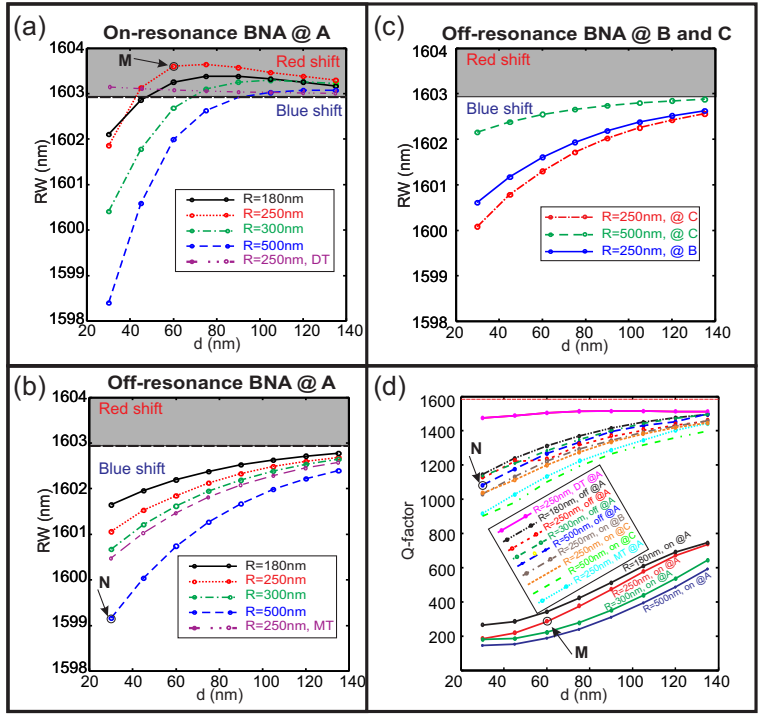}
\caption{(a) Resonance wavelengths (RW)  calculated in the case of a resonant BNAT placed above the CL7 center in position A (see white arrow in figure \ref{cl7}d) as a function of the tip-to-CL7 tip $d$. The RW shift is both positive and negative with respect to $d$ demonstrating the double-way tunability. The case of a pure dielectric tip (similar to that of ref. \cite{burresi:prl10}) is also presented in dashed-dotted-dotted magenta line. (b) same as (a) in the case of non-resonant BNAT. Electric effect is very weak and only blue shifts are obtained. The case of a pure metallic tip (as in ref. \cite{maksymov:oe11}) is presented for comparison. (c) Same as (a) and (b) but for BNAT apex positioned at the vertical of B and C positions (see figures \ref{cl7}c and d). (d) Q-factors of the thirteen studied configurations corresponding to different apex radii (from $R=180nm$ to $R=500nm$) and BNA state (on or off resonance) including uncoated dielectric tip (DT) and metallic tip (MT) as a function of the BNAT-to-CL7 distance $d$. The red dashed horizontal lines at the top of the figure corresponds to the case of the unperturbed CL7 cavity ($Q_{CL7}=1590$).} \label{shifts}
\end{figure}
The electric term of Eq. \ref{relativeshift} can be determined by subtracting the pure magnetic effect of figure \ref{cpcp}a, observed in the case of off-resonant BNA, from the effect corresponding to a resonantly excited BNA (keeping the tip at the same position and rotating the BNA by $90^o$). The obtained values are presented in figure \ref{cpcp}b. Contrarily to the magnetic term, the relative shift (in logarithmic scale) of the induced electric dipole exhibits a non-linear behavior versus the BNAT-to-PC distance clearly denoting the non validity of the quasi-static approximation. A maximum of red-shift is obtained (see vertical dashed lines in figure \ref{cpcp}b) corresponding to a critical coupling between the two resonators (i.e. maximum energy transfer between them) \cite{belarouci:oex10}. As expected, this critical tip-to-PC distance increases with the tip apex radius due to the redistribution of the electromagnetic energy inside and around the cavity \cite{lefeber:natphot14}.

Other simulations were done for different configurations including both the modification of the tip position, the BNA direction (on- and off-resonance) and the tip apex radius. These configurations are compared to the case of an uncoated dielectric tip (DT) and to the case of a metallic tip (MT). The numerical results are
presented in figure \ref{shifts}. Because negative and positive shifts occur, we plotted here the RW values (figure
\ref{shifts}a-c) instead of the relative shift as in figure \ref{cpcp}. The Q-factor variations are also given in figure \ref{shifts}d. As expected, figure \ref{shifts}a shows the competition effect between the induced electric and the magnetic dipoles at the tip apex. This competition only appears when the BNA resonates (position A given by white arrow on figure \ref{cl7}d) i.e. when the antenna gap is immersed in a transverse electric field directed along its metallic arm (y-direction here). If the BNA position corresponds to a node of transverse electric field (positions B or C pointed by the green and the blue arrows respectively) or if it is off-resonance (even in the position A), only resonance blueshifts occur. In addition, for BNAT with large apex radius, the blue shift is
more pronounced meaning larger magnetic polarizability of the tip apex as already demonstrated above. Meanwhile, the quality factor given in figure \ref{shifts}d is significantly
reduced when the BNA resonates because the stored electromagnetic energy that
was confined in the CL7 cavity is now dissipated by the BNA. This leads
to a decrease of the photon lifetime and thus to a decrease of the
Q-factor of the cavity resonance. Similar decreasing of Q-factor was also obtained in refs. \cite{koenderink:prl05,vignolini:prl10} even if no electric resonance is excited (passive dielectric tip).

Note that, on one hand, for a big tip with $R=500~nm$ (blue dashed line in figure \ref{shifts}a), the two
effects (blue and red shifts) compensate for a distance $d=90~nm$
while this occurs at smaller distance ($d\simeq40~nm$) for a tip apex radius of $R=250~nm$ (red dotted line in figure \ref{shifts}a). Nevertheless, for both configurations, the quality
factor of the whole structure decreases to
$Q_{90}^{500}=310$ and $Q_{40}^{250}=210$ respectively instead of
$Q_{CL7}=1590$ for the unperturbed cavity (see figure \ref{shifts}d). Consequently, from ref. \cite{mivelle:oe14}, we estimate that the energy dissipated
by the BNAT (radiated inside the fiber and/or absorbed by the metal)
is around $80\%$ and $86\%$ of the total energy
initially stored in the cavity respectively. On
the other hand, for $R=500~nm$, a $\Delta\lambda=5~nm$ blueshift is
achieved by only approaching the BNAT at $30~nm$ from the PC. Larger shifts can occur for smaller tip-to-PC distances as already experimentally demonstrated in ref. \cite{grosjean:apl12} where 16nm blueshift were observed for $d=10nm$. This value corresponds to an efficient tunability of the cavity
resonance in comparison with the value of $\Delta \lambda=1.7~nm$ that was
recently obtained by focusing effect on plasmonic nanorod coupled
with a PC cavity \cite{maksymov:oe11}.

\begin{figure}[ht]
\centering
\includegraphics[width=10cm]{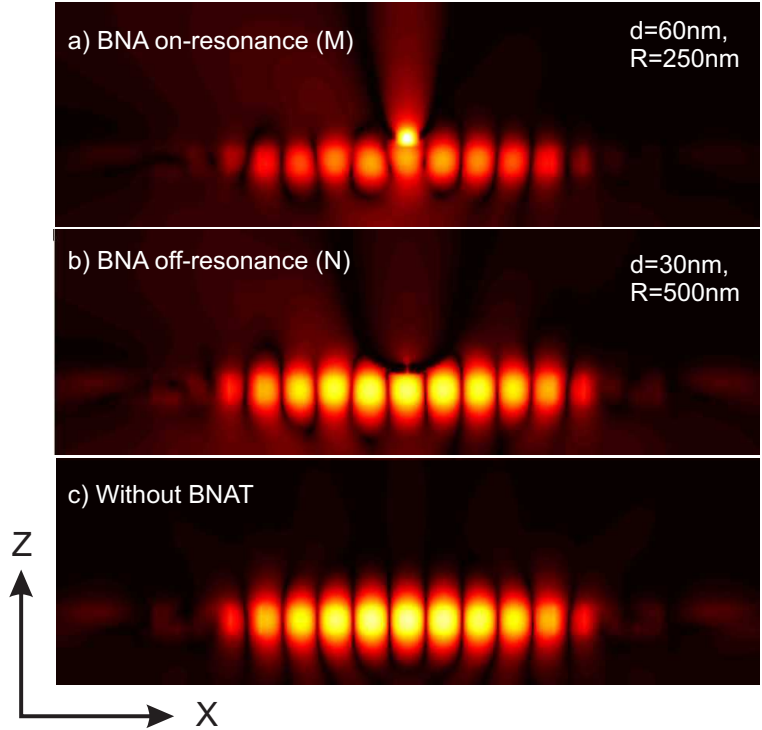}
\caption{(a-c) calculated electric intensity distributions (in
logarithmic scale) corresponding to the case of the two points M  (in figure \ref{shifts}a and d) and
N (in figure \ref{shifts}b and d) in addition to the case of the unperturbed PhC
respectively. One can clearly see in (a) that the BNA pumps the PhC
energy and then dissipates it mainly inside the tip while the spatial extension of the cavity mode is
clearly reduced in (b), compared to the unperturbed cavity in (c), due to the presence of the metal tip apex that squeezes the electromagnetic evanescent field down to the cavity. This result is consistent with the obtained discrepancy between the Q-factor of these two
configurations given in figure \ref{shifts}b} \label{distYZ}
\end{figure}

Finally, the light distribution inside the structure is shown in fig. \ref{distYZ}
at the resonance wavelength of the hybrid structure for two contrasting configurations
indicated by the points $M$ and $N$ in figure \ref{shifts}d. Our main
goal is to point out the spatial expansion or squeezing of the mode
with regard to the case of the unperturbed cavity resonance given in
figure \ref{distYZ}c. As it can be seen on figure \ref{distYZ}a, the
resonance of the BNA leads to a dramatic reduction of the light
energy inside the CL7 due to a significant penetration inside the
gap zone (compared to the figure \ref{distYZ}c). On the contrary, a
volume reduction of the energy distribution occurs (see figure
\ref{distYZ}b) near the BNAT accompanied with a small decrease of the
photon lifetime in the CL7 cavity (slight reduction of the resonance
quality factor) when the BNA is off-resonant.

This study clearly demonstrated that a BNAT tip has the ability of
red- and blue-shifting the resonance of an optical nanocavity in a fine
and reversible post-production tuning. Nevertheless, as in ref. \cite{koenderink:prl05}, redshifts are
achieved at the cost of significant Q-factor reduction. An
experimental demonstration of the coupling between the same PC
cavity (CL7) and a BNAT was already performed \cite{mivelle:oe14}
but only blueshift of the cavity resonance was observed because the
nano-antenna effect was not as strong as predicted. This points out
that the tip design is critical in the achievement of the desired
effect onto the nanocavity. Based on this configuration, alternative
tip and antenna geometries may be found to accomplish this two-way
spectral tunability of a nanocavity resonance while preserving its
Q-factor even in the case of redshift by inhibiting the far-field scattering toward the antenna.

\textbf{Acknowledgements:} The authors are indebted to Prof. K.
Kuipers for helpful discussions and to Dr. M. Dankar for her
technical advise. This work is funded by ``Agence Nationale de la Recherche" under contract number
ANR10-NANO-002, the program ``BQR PRES Bourgogne Franche-Comté''.  It is supported by the ``P\^ole de
comp\'etitivit\'e Microtechnique'', the ``Labex ACTION'' program (Contract ANR-11-LABX-01-01) and the ``M\'esocentre''
of the University of Franche-Comt\'e.

\end{document}